\documentclass[prb,twocolumn,showpacs,preprintnumbers]{revtex4}
\usepackage{graphics}
\usepackage{graphicx}% Include figure files
\usepackage{dcolumn} % Align table columns on decimal point
\usepackage{bm}
\usepackage{float}
\usepackage{epsfig}
\begin{document}
\title{Origin of ferromagnetism in Cs$_2$AgF$_4$: importance of Ag - F 
covalency}
\author{Deepa Kasinathan,$^1$ A. B. Kyker,$^1$ and D. J. Singh$^2$}
\affiliation{$^1$Department of Physics, University of California Davis,
  Davis, CA 95616}
\affiliation{$^2$Condensed Matter Sciences Division,
Oak Ridge National Laboratory, Oak Ridge,TN 37831-6032}
\date{\today}

\begin{abstract}

The magnetic nature of Cs$_{2}$AgF$_{4}$,
an isoelectronic and isostructural analogue of 
La$_2$CuO$_4$,
is analyzed using density functional calculations.
The ground 
state is found to be
ferromagnetic and nearly half metallic.
We find
strong hybridization
of Ag-$d$ and F-$p$ states.  Substantial moments 
reside on the F atoms,
which is unusual for the halides and reflects the chemistry of the
Ag(II) ions in this compound. This provides the mechanism
for ferromagnetism, which we find to be itinerant in character,
a result of a Stoner instability enhanced by Hund's coupling on the
F.

\end{abstract}

\maketitle

% \section{Introduction}

Cs$_2$AgF$_4$ is a member of a family of Ag(II) fluorides that form
in perovskite and layered perovskite structures. The distinguishing
feature is the presence of Ag(II), which is a powerful oxidizing agent.
\cite{hoppe1,hoffmann}
This compound  was first synthesized in 1974 by
Odenthal and co-workers. \cite{hoppe}
It occurs in the tetragonal K$_{2}$NiF$_{4}$ layered perovskite
structure.
This is the same structure as the parent of the high temperature
superconducting cuprates, La$_2$CuO$_4$. 
Cs$_2$AgF$_4$ shows no tilts or rotations of the octahedra,
which are common in oxide layered perovskites.
Synthesis of isostructural Na$_2$AgF$_4$ and K$_2$AgF$_4$ was also
reported and these compounds also have the 
K$_{2}$NiF$_{4}$ structure. All three compounds are reported as being
blue or purple in appearance
and ferromagnetic. While transport measurements have
not been reported for these compounds, it is known that the related
distorted perovskite compound KAgF$_3$ is metallic at high temperatures,
and then has a metal insulator transition coincident with an antiferromagnetic
ordering temperature. \cite{g-kagf3}

In the doped high-T$_c$ cuprates, superconductivity develops from
a paramagnetic metallic phase, with Fermi surfaces coming from
hybridized Cu $d$ - O $p$ bands. These are formally antibonding bands of
$d_{x^2-y^2}$ - $p_\sigma$ character. \cite{pickett}
While the theory of high temperature
cuprate superconductivity remains to be established, it is widely held
that the phenomenon is associated with the physics of the undoped
compounds, which are antiferromagnetic Mott insulators.
Specifically,
it is thought that there is a relationship between superconductivity
and the antiferromagnetic
fluctuations associated with the correlated $d$ electrons of cuprates.
Cs$_2$AgF$_4$ has interesting similarities to the high-T$_c$ cuprates.
As mentioned, it is isostructural, featuring AgF$_2$ sheets in place
of CuO$_2$ sheets, it has a transition element with a $d^9$ configuration,
and it is magnetic. Moreover, related compounds have been shown
both in band structure calculations and X-ray
photoelectron spectroscopy experiments to
display significant Ag - F covalency, reminiscent of the Cu - O hybridization
in the cuprates. \cite{hoffmann,jaron,g2}
These similarities and other considerations have led to speculations
about possible high temperature superconductivity in Ag(II) fluorides.
\cite{hoffmann,g3}
One puzzling difference between the cuprates and the layered Ag(II)
fluorides is that the undoped cuprates are antiferromagnetic, while
the argentates are ferromagnetic.
One possible explanation would be an orbital ordering that favors
ferromagnetism within a superexchange
framework, as was recently suggested.
However, neutron measurements did not detect the symmetry lowering that
would occur in this case. \cite{mclain}

Here we use electronic structure calculations to elucidate the
electronic structure of Cs$_2$AgF$_4$ and the origin of its magnetic
properties.
A previous density functional calculation for
this material found it to be a covalent metal,\cite{hoffmann}
with a substantial density of states (DOS)
at the Fermi level (E$_{F}$) in the absence of magnetism.

% \section{Approach}

We did electronic structure calculations within the local spin density
approximation (LSDA) and the generalized gradient approximation (GGA),
\cite{pw,pbe}
using the general potential linearized augmented planewave method,
with local orbitals,
\cite{lapw,lo} as implemented in the WIEN2K program. \cite{wien}
The augmented planewave plus local orbital extension was used for
the Ag $d$ and semicore levels.
\cite{apw}
The valence states were treated in a scalar relativistic approximation,
while the core states were treated relativistically.
Well converged basis set sizes and Brillouin zone samplings were
employed. Except as noted otherwise, the LAPW sphere radii were
2.0 $a_0$ and 1.85 $a_0$ for the metal and fluorine atoms, respectively.
The basis set cut-off was chosen to be $RK_{max}$=7.0, where $R$ is the
radius of the F sphere.
We tested the convergence by comparison of LSDA results with
an independent code, employing the LAPW augmentation with local orbitals
and with higher basis set cut-offs as well as different sphere radii.

The
structural data were obtained from the
report\cite{hoppe} of Odenthal and co-workers:
$a$ = 4.58\AA, $c$ = 14.19\AA,
including the two internal parameters corresponding to the Cs and apical
F heights above the AgF$_2$ square planar sheets.
Minimization of the forces in the LDA approximation yielded a value of 
z$_{\rm Cs}$=0.361 and z$_{\rm F}$=0.147, in close
agreement with the experimental values of z$_{\rm Cs}$=0.36 and
z$_{\rm F}$=0.15.

% \section{Electronic Structure and Magnetism}

Within the LSDA we find a Cs$_2$AgF$_4$ to be a metal on the
borderline of ferromagnetism. Fixed spin moment calculations
showed a non-spin-polarized ground state, but with a 1 $\mu_B$ per
formula unit fully polarized solution only 35 meV higher in energy.
We also did LSDA calculations applying fields only inside the Ag
LAPW spheres, which were chosen to be 2.1 $a_0$ in radius for
this purpose. With 5 mRy fields of this type in a ferromagnetic pattern,
moments of 0.35 $\mu_B$ were induced in the Ag spheres, and moments
also appeared in the F spheres, for a total spin magnetization of
0.62 $\mu_B$. Application of the same field in an in-plane $c$(2x2)
antiferromagnetic pattern yielded induced Ag moments in the spheres of only
0.17 $\mu_B$, with a small moment also appearing on the apical F,
but no moments on the in-plane F, as is required by symmetry.
This shows the system to be much closer to ferromagnetism than
antiferromagnetism at the LSDA level, and suggests an important
role for the in-plane F in the magnetism.

\begin{figure}
\epsfig{angle=270,width=0.90\columnwidth,file=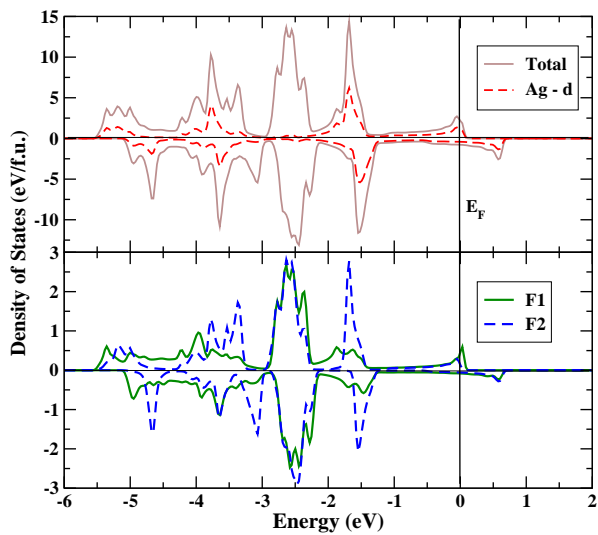}
\caption{(color online) Calculated DOS of ferromagnetic Cs$_{2}$AgF$_{4}$. 
{\it Top Panel} : The total (continuous/brown) and the orbitally
resolved $4d$ states (dashed/red)
of Ag, via projection onto the LAPW sphere.
{\it Bottom Panel}: The orbitally resolved $2p$ states of the two types of
Fluorine, 
F1 (continuous/green) and F2 (dashed/blue).
Note the strong mixture of F and Ag states in the valence bands.
}
\label{dos}
\end{figure}

\begin{figure}[b]
\epsfig{width=0.90\columnwidth,angle=0,file=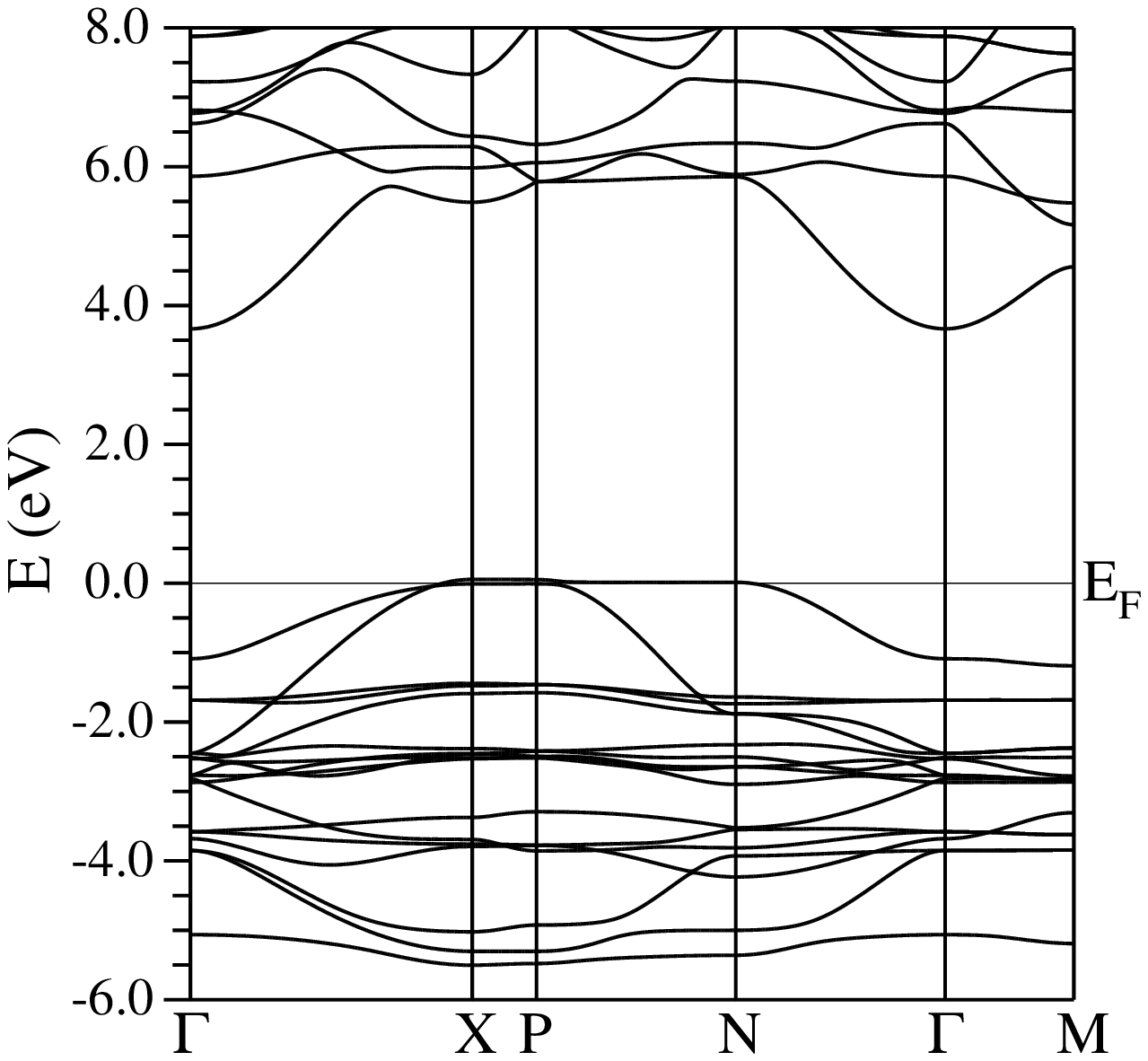}
\vspace{0.3cm}
\epsfig{width=0.90\columnwidth,angle=0,file=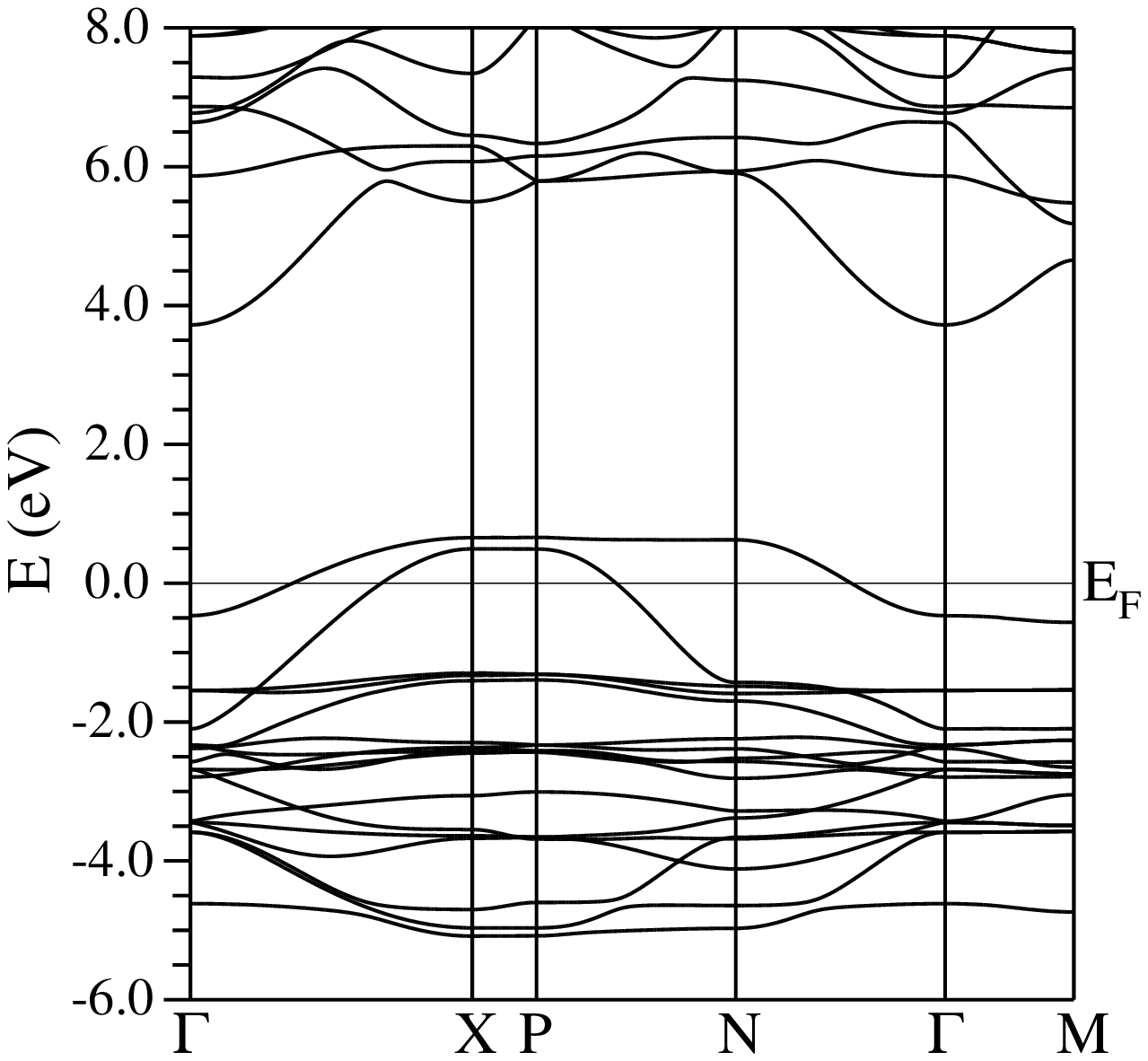}
\caption{GGA band structure of ferromagnetic Cs$_{2}$AgF$_{4}$ using GGA,
for majority (top) and minority (bottom) spins. Note the
near half metallic character and the large gap between
the hybridized Ag $d$ - F $p$ derived valence bands and the unoccupied
Cs derived bands above.}
\label{bands}
\end{figure}

\begin{figure}[b]
\epsfig{width=0.90\columnwidth,angle=0,file=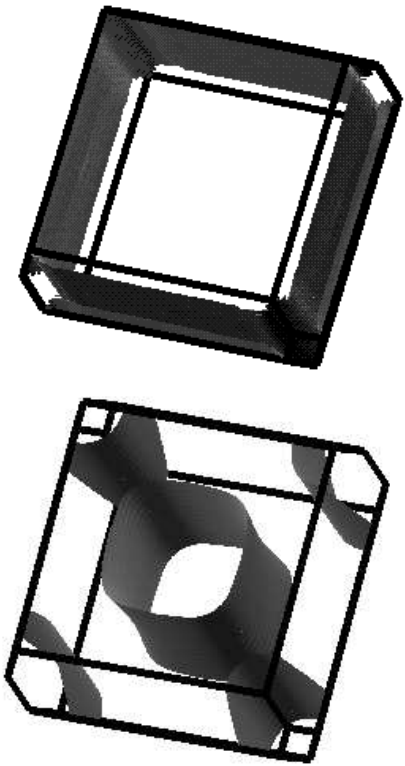}
\caption{GGA majority (top)
and minority (bottom) Fermi surfaces of ferromagnetic Cs$_{2}$AgF$_{4}$.
}
\label{fermi}
\end{figure}

Within the GGA, we obtain a ferromagnetic ground state, with spin
magnetization, $M=0.9 \mu_B$ and energy 6 meV below the non-spin
polarized solution.
However, we do not find any metastable antiferromagnetic solution, implying
itinerant magnetism, in particular,
the absence of stable local moments.
The calculated electronic density of states (DOS) for the
ferromagnetic ground state is
shown in Fig. \ref{dos}. The band structure is shown in Fig. \ref{bands},
and the Fermi surface in Fig. \ref{fermi}.
The band structure is expected to be two dimensional, due to the
bonding topology, which has 180$^\circ$ Ag-F-Ag links in the AgF$_2$
sheets, but no direct Ag-F-Ag connections in the $c$-axis direction.
This in fact is the case. \cite{disp-note}
As may be seen, Cs$_2$AgF$_4$ is close to a half metal,
with the Fermi energy being near a band edge in the majority channel, but
not in the minority channel.
The minority spin Fermi surface consists of small hole cylinders
running along the zone corner (from the $d_{x^2-y^2}$ band) and electron
cylinders around the zone center i(from the $d_{z^2}$ band).
The majority spin Fermi surface consists of a single large square
cylindrical electron surface that almost fills the Brillouin zone,
leaving a small region of holes around the zone boundary.

Cs$_{2}$AgF$_{4}$ has two type of F sites forming
distorted Ag centered octahedra; one is 
in the AgF$_2$ sheets
(referred as F1 in this paper), and the other is the
apical F along the $c$ - axis 
(referred as F2 in this paper).
The apical Ag - F2 distance is slightly smaller than the in-plane Ag - F1
distance.
A key point is that the F1 atoms bridge the Ag atoms in the sheets, with
180$^\circ$ bonds, while the apical, F2 atoms connect to only one
Ag atom and therefore are not bridging.

Examining the DOS and projections in more detail, one may note
that the valence bands have substantially mixed Ag $d$ - F $p$ character,
especially near the bottom and top of the manifold where $e_g$ - $p_\sigma$
bonding and antibonding combinations occur. This hybridization involves both
F1 and F2 atoms, and is consistent with previous results for Ag(II) fluorides.
\cite{hoffmann,jaron,g2}
The result is a very stable metallic electronic structure, with substantial
F character at the Fermi energy, $E_F$,
and a valence band width of $\sim$ 5.5 eV.
This strong hybridization can be understood in
chemical terms considering the very strongly oxidizing character of Ag(II),
which in this compound partially oxidizes F$^{-}$.
Thus covalency in this compound is a consequence of the unusual valence
state of Ag.
Turning to the band structure, there are two bands crossing $E_F$ in
the minority spin channel. These are the $d_{x^2-y^2}$, which hybridizes
with the in-plane F, and the $d_{z^2}$ hybridized with the apical F. 
The $d_{x^2-y^2}$ - F1 combination has greater dispersion because of the
in-plane topology, mentioned above. However, the $d_{z^2}$ - F2 combination
is higher lying, with the result that the two band maxima
nearly coincide. The higher lying position of the $d_{z^2}$ - F2
is readily explained by the fact that these bands are
antibonding $e_g$ - $p_\sigma$ and the Ag - F2 bond is shorter.
In the minority spin channel, the $d_{z^2}$ band extends from
from -0.5 to 0.7 eV (relative to $E_F$),
while the $d_{x^2-y^2}$ band extends from -2.1 to 0.5 eV.
In the absence of the lighter $d_{x^2-y^2}$ band, one would have a
half-filled $d_{z^2}$ band. Because the $d_{x^2-y^2}$ is in fact present,
the $d_{z^2}$ is slightly electron doped away from half filling. This is
in contrast to the cuprates, where only a $d_{x^2-y^2}$ band is active,
and this band is hole doped away from half-filling in the highest $T_c$
compounds.

The small size of F$^-$ relative to O$^{2-}$ emphasizes the effect of
the perovskite bonding topology in the band structure. This is because
direct F - F hopping is reduced by its small size, relative to O in oxides,
and the strongly oxidizing nature of F and Ag(II) push the Cs conduction
bands to high energy, reducing the assisted hopping via Cs for the valence
bands. Thus, the hopping is dominated by nearest neighbor Ag-F channels,
so for example, the $d_{xz} - p_\pi$ and $d_{yz} - p_\pi$
derived bands take strong
one dimensional character and are seen to be almost perfectly flat along
some directions as seen in Fig. \ref{bands}.

The mixed character of the bands is reflected in the distribution of the
magnetic moments in the ferromagnetic ground state.
Of the total spin moment of 0.9 $\mu_{B}$, only 0.5 $\mu_{B}$ lies
within the Ag LAPW sphere, radius 2.0 $a_0$. The remaining $\sim$ 40\%
of the magnetization is F derived, approximately equally divided between
the F1 and F2 sites.
This is important for understanding the itinerant ferromagnetic ground
state that we find.
First of all, the large moments on the in-plane F1 atoms seen both in
the GGA ferromagnetic ground state and in the LSDA calculations with
ferromagnetic fields in the Ag (but not the F) spheres, mean that
there is a contribution to the energy from the F polarization.
F$^-$ is a relatively small anion, at the end of the first row of the
periodic table. Thus, when magnetic, it can have a strong Hund's coupling.
This provides a generalized double exchange mechanism for favoring
ferromagnetism, similar to the mechanism in SrRuO$_3$. \cite{singh-ru,mazin-ru}
In the ferromagnetic case, the F1 atoms take moments due to the
hybridized character of the bands around $E_F$ and contribute to the
Stoner instability through their Hund's coupling. With antiferromagnetic
ordering, no induced moments can be present on the F1 atoms by symmetry,
and therefore the Hund's coupling on these sites cannot stabilize the
magnetism. Thus,
the fact that the moments become unstable in an in-plane antiferromagnetic
configuration supports this picture.
Different from the ruthenates, the hybridized states in Cs$_2$AgF$_4$
involve $e_g$ instead of $t_{2g}$ states, and the F$^{-}$ ion is
much smaller than O$^{2-}$.

We studied the stability of this ferromagnetic, two-band electronic structure,
using LDA+U calculations and treating the Coulomb $U$ as a parameter.
We found, as expected, that a local moment, insulating state could
be obtained. However this
only happened when using a very high value $U$=7 eV. This is an
unreasonably large value for a 4$d$ ion in a screening environment. 
The reason for the weak effect of more realistic values of $U$
is that the bands are strongly hybridized, and are really mixed F $p$ -
Ag $d$ bands, and not narrow bands built from the Ag $d$ orbitals.
Thus we conclude that on-site Coulomb correlations do not have
a large effect on the electronic structure or magnetism of this compound.
This is in contrast to the undoped cuprates, where the LSDA and GGA
approximations incorrectly predict non-magnetic metallic ground states,
and the Hubbard $U$ is crucial for obtaining moment formation and an
insulating ground state.

% \section{Summary and Conclusions}

To summarize,
density functional calculations of the electronic structure of 
Cs$_2$AgF$_4$ show strong covalency between Ag $d$ and F $p$
states. Within the GGA, the ground state is ferromagnetic,
and is stabilized by Hund's coupling on the in-plane F atoms
which occurs due to F participation
in the magnetism resulting from the $e_g$ - $p_\sigma$ hybridization.
The electronic structure is nearly half-metallic, and not insulating.
It would be of interest to experimentally test
the prediction of a metallic electronic structure.

The resulting picture of the electronic structure and magnetism is
very different from the undoped cuprates. (1) Cs$_2$AgF$_4$ has
two active bands: $d_{x^2-y^2}$ and $d_{z^2}$; neither is exactly
half-filled; (2) moment formation in Cs$_2$AgF$_4$ is due to a
Stoner type mechanism as opposed to on-site Coulomb
repulsions that are crucial in the cuprates; (3) the magnetism
has strong itinerant character due to F participation, as opposed
to the local moment superexchange mediated character of cuprate
antiferromagnetism; and (4) we find ferromagnetism with the ideal
tetragonal structure; orbital ordering to obtain ferromagnetism
within a superexchange mediated framework is not needed.
We note that the predicted F contributions to the magnetism
are large enough to be detected using neutron scattering.

Finally, we note that the mechanism for ferromagnetism in Cs$_2$AgF$_4$
is quite robust, and would expected to occur in other Ag(II) fluorides
with similar bond lengths and topologies. Since it does not rely on
small structural effects, it provides a ready explanation for the
observed ferromagnetism in the other $A_2$AgF$_4$ compounds. Furthermore,
the above picture of itinerant
magnetism may be more generally applicable to other Ag(II) fluoride compounds.
For example, KAgF$_3$ shows a metal insulator transition
coincident with an antiferromagnetic ordering. \cite{g-kagf3}
This is much
more natural in an itinerant system than in a strongly correlated
local moment system, which would tend to be insulating on both sides
of the ordering temperature at odd integer band fillings. The
structure of that compound shows compressed octahedra and 
Ag-F-Ag chains along the $c$-axis direction with short bond lengths.
Assuming that the above mechanism applies
also to this compound, one may expect ferromagnetic chains along
$c$. Considering that the ground state is known to be antiferromagnetic,
one may anticipate a C-type ordering of antiferromagnetic $a-b$ planes,
stacked ferromagnetically in that case. In any case,
in perovskite derived Ag(II) fluorides, the mechanism
that we propose would generally favor ferromagnetism or complex
antiferromagnetic states, with ferromagnetic interactions along some
bonding directions, as opposed to simple G-type ordering.

\begin{acknowledgments}

We are grateful for helpful discussions with W.E. Pickett and J. Turner.
Research at ORNL was
sponsored by the Division of Materials Sciences and Engineering,
Office of Basic Energy Sciences, U.S. Department of Energy,
under contract DE-AC05-00OR22725 with Oak Ridge National Laboratory,
managed and operated by UT-Battelle, LLC.
Work at UC Davis was supported by DOE contract DE-FG03-01ER45876.

\end{acknowledgments}


\begin{thebibliography}{100}

\bibitem{hoppe1}
R. Hoppe, Angew. Chem. Int. Ed. Engl. {\bf 20}, 63 (1981).

\bibitem{hoffmann}
W. Grochala and R. Hoffmann, Angew. Chem. Int. Ed. Engl. {\bf 40}, 2742 (2001).

\bibitem{hoppe}
R.-H. Odenthal, D. Paus and R. Hoppe,
Z. Anorg. Allg. Chem. {\bf 407}, 144 (1974).

\bibitem{g-kagf3}
W. Grochala and P.P. Edwards,
Phys. State Sol. (b) {\bf 240}, R11 (2003).

\bibitem{pickett}
W.E. Pickett, H. Krakauer, R.E. Cohen and D.J. Singh,
Science {\bf 255}, 46 (1992).

\bibitem{jaron}
T. Jaron, W. Grochala and R. Hoffmann, 
Phys. State Sol. (b) {\bf 242}, R1 (2005).

\bibitem{g2}
W. Grochala, R.G. Egdell, P.P. Edwards, Z. Mazej and B. Zemva,
ChemPhysChem {\bf 4}, 997 (2003).

\bibitem{g3}
W. Grochala, A. Porch and P.P. Edwards,
Solid State Commun. {\bf 130}, 137 (2004).

\bibitem{mclain}
S.E. McLain, D. A. Tennant, J. F. C. Turner, T. Barnes, M. R. Dolgos,
Th. Proffen, B. C. Sales and R. I. Bewley,
cond-mat/0509194 (2005).

\bibitem{pw}
J.P. Perdew and Y. Wang, Phys. Rev. B {\bf45}, 13244 (1992).

\bibitem{pbe}
J.P. Perdew, K. Burke and M. Ernzerhof, Phys. Rev. Lett. {\bf 77}, 3865 (1996).

\bibitem{lapw}
D.J. Singh and L. Nordstrom, Planewaves, Pseudopotentials and the LAPW
Method, 2nd Ed. (Springer, Berlin, 2006).

\bibitem{lo}
D. Singh, Phys. Rev. B {\bf 43}, 6388 (1991).

\bibitem{wien}
P. Blaha, K. Schwarz, G. K. H. Madsen, D. Kvasnicka, and J. Luitz,
{\small WIEN2K}, 
2002, an Augmented Plane Wave + Local Orbitals Program for calculating crystal
properties (Karlheinz
Schwarz, Technische Universitat Wien, Austria).

\bibitem{apw}
E. Sjostedt, L. Nordstrom and D.J. Singh, Solid State Commun. {\bf 114},
15 (2000).

\bibitem{disp-note}
This can be seen for the valence
bands in Fig. \ref{bands} from the $\Gamma -M$ and 
$X - P$ lines,
which are along $k_z$ in the body centered tetragonal
zone. The two-dimensional character is also evident in the Fermi surfaces.
The higher lying (mainly Cs) derived states above the
gap are more three dimensional in character.

\bibitem{singh-ru}
D.J. Singh, J. Appl. Phys. {\bf 79}, 4818 (1996).

\bibitem{mazin-ru}
I.I. Mazin and D.J. Singh, Phys. Rev. B {\bf 56}, 2556 (1997).

\end{thebibliography}
\end{document}